\documentclass[%
 aip,
 amsmath,amssymb,
reprint,%
]{revtex4-1}

\usepackage{isomath}
\usepackage{amsmath,amsthm}
\usepackage{amsbsy}
\usepackage{amssymb}
\usepackage{amscd}
\usepackage{amsfonts}
\usepackage{stmaryrd}
\usepackage{siunitx}
\usepackage{euscript}
\usepackage[utf8]{inputenc}
\usepackage[T1]{fontenc}
\usepackage{newtxtext} 
\everymath{\displaystyle}
\usepackage{exscale}

\usepackage{graphicx}
\usepackage{boxedminipage}
\usepackage{calc}
\usepackage[usenames,dvipsnames]{xcolor}
\graphicspath{ {media/} }

\usepackage{setspace}
\usepackage{enumitem}
\setitemize{noitemsep,topsep=0pt,parsep=0pt,partopsep=0pt}
\setenumerate{noitemsep,topsep=0pt,parsep=0pt,partopsep=0pt}
\setdescription{noitemsep,topsep=0pt,parsep=0pt,partopsep=0pt}

\usepackage[colorinlistoftodos, color=green!40,prependcaption]{todonotes}

\usepackage{soul} 

\usepackage{xr-hyper}
\makeatletter
\newcommand*{\addFileDependency}[1]{
  \typeout{(#1)}
  \@addtofilelist{#1}
  \IfFileExists{#1}{}{\typeout{No file #1.}}
}
\makeatother

\usepackage[,colorlinks,urlcolor=blue,citecolor=black]{hyperref}
\usepackage[normalem]{ulem}

\setuptodonotes{inline}
\usepackage{isomath}
\usepackage{amsmath}
\usepackage{amssymb}
\usepackage{amscd}
\usepackage{amsfonts}

\newcommand{\bfmu}{\mathbold {\mu}}

\newcommand{\parderiv}[2]{\frac{\partial #1}{\partial #2}}

\newcommand{\bfx}{{\mathbold x}}

\usepackage{dcolumn}
\usepackage{caption}
\usepackage{subcaption} 

\usepackage{csquotes}
\usepackage{svg}
\usepackage[symbol]{footmisc}

\usepackage{scalerel}
\usepackage{tikz}
\usetikzlibrary{svg.path}

\definecolor{orcidlogocol}{HTML}{A6CE39}
\tikzset{
  orcidlogo/.pic={
    \fill[orcidlogocol] svg{M256,128c0,70.7-57.3,128-128,128C57.3,256,0,198.7,0,128C0,57.3,57.3,0,128,0C198.7,0,256,57.3,256,128z};
    \fill[white] svg{M86.3,186.2H70.9V79.1h15.4v48.4V186.2z}
                 svg{M108.9,79.1h41.6c39.6,0,57,28.3,57,53.6c0,27.5-21.5,53.6-56.8,53.6h-41.8V79.1z M124.3,172.4h24.5c34.9,0,42.9-26.5,42.9-39.7c0-21.5-13.7-39.7-43.7-39.7h-23.7V172.4z}
                 svg{M88.7,56.8c0,5.5-4.5,10.1-10.1,10.1c-5.6,0-10.1-4.6-10.1-10.1c0-5.6,4.5-10.1,10.1-10.1C84.2,46.7,88.7,51.3,88.7,56.8z};
  }
}

\newcommand\orcidicon[1]{\href{https://orcid.org/#1}{\mbox{\scalerel*{
\begin{tikzpicture}[yscale=-1,transform shape]
\pic{orcidlogo};
\end{tikzpicture}
}{|}}}}

\usepackage{verbatim}

\newcommand{\etal}{\textit{et al.}}

\makeatletter
\def\@email#1#2{%
 \endgroup
 \patchcmd{\titleblock@produce}
  {\frontmatter@RRAPformat}
  {\frontmatter@RRAPformat{\produce@RRAP{*#1\href{mailto:#2}{#2}}}\frontmatter@RRAPformat}
  {}{}
}%
\makeatother

\begin{document}

\preprint{AIP/123-QED}

\title[Temporal Metasurfaces]{An Energy Conserving Mechanism for Temporal Metasurfaces} 
\affiliation{Department of Mathematics, University of Utah, Salt Lake City, UT 84112, U.S.A}
\author{Kshiteej J. Deshmukh\orcidicon{0000-0002-6825-4280}\thanks{*}}

\author{Graeme W. Milton\orcidicon{0000-0002-4000-3375}}%
\email[Corresponding author:\ ]{kjdeshmu@math.utah.edu}



\date{\today}
\begin{abstract}
Changing the microstructure properties of a space-time metamaterial while a wave is propagating through it, in general requires addition or removal of energy, which can be of exponential form depending on the type of modulation. 
This limits the realization and application of space-time metamaterials. 
We resolve this issue by introducing a mechanism of conserving energy at temporal metasurfaces in a non-linear setting.
The idea is first demonstrated by considering a wave-packet propagating in a discrete medium of 1-d chain of springs and masses, where using our energy conserving mechanism we show that the spring stiffness can be incremented at several time interfaces and the energy will still be conserved. 
We then consider an interesting application of time-reversed imaging in 1-d and 2-d spring-mass systems with a wave packet traveling in the homogenized regime.
Our numerical simulations show that, in 1-d, when the wave packet hits the time-interface two sets of waves are generated, one traveling forward in time and the other traveling backward. 
The time-reversed waves re-converge at the location of the source and we observe its regeneration.  
In 2-d, we use more complicated initial shapes and, even then, we observe regeneration of the original image or source. 
Thus, we achieve time-reversed imaging with conservation of energy in a non-linear system.
The energy conserving mechanism can be easily extended to continuum media.
Some possible ideas and concerns in experimental realization of space-time media are highlighted in conclusion and in the supplementary information.
\end{abstract}
\maketitle


The spatio-temporal modulation of metamaterial microstructure has led to the discovery of novel and exciting properties like space-time mirrors\cite{Bacot2016}, chromatic birefringence\cite{Akbarzadeh2018}, space-time cloaking\cite{mccall2010spacetime}, and photonic-time-crystals (PTCs)\cite{zeng2017photonic}. 
Their added flexibility in manipulating electromagnetic, acoustic, and elastic wave propagation has led to an emerging widespread interest in space-time metamaterials.
These materials have an immense potential for applications in radio and optics for electromagnetic cloaking, frequency multiplication, bandwidth-matching, and filtering. 
A common feature of most space-time media is the nonreciprocal scattering\cite{taravati2017nonreciprocal, caloz2018electromagnetic} of waves which can be potentially applied to realize magnetless nonreciprocal devices.
A mechanism of controlling photon flow using effective magnetic fields obtained by dynamic modulation was shown by Fang \etal \cite{fang2012realizing}.
Introducing a \textit{synthetic frequency} dimension obtained by time modulation of refractive index has led to interesting ideas of simulating Weyl points and nodal loops, and 4-dimensional quantum Hall physics\cite{Boada2012, Celi2014, ozawa2016synthetic, zhang2016simulating}.
A thorough summary of the research on space-time  metamaterials until the year $2020$ is provided in a two part paper by Caloz \etal\cite{caloz2019spacetime, Caloz2020}, and a review of all the work on this topic can be found in the excellent book by Lurie\cite{lurie2007introduction}.

Space-time metamaterials, also called \textit{dynamic metamaterials}, can have time, or space-time interfaces at which there is a discontinuous change in their properties.
Some of the models for space-time microstructures are space-time laminates, space-time checkerboards, and electrical transmission line structures\cite{Cassedy1967, Lurie2016} with the capacitance or inductance being temporally modulated.  
In simple space-time laminates and checkerboards the microstructure properties can be doubly periodic in space and time with piecewise constant values changing abruptly at space-time interfaces. 
A recent application of space-time materials is the idea of an inverse prism\cite{Akbarzadeh2018}.
Conventional prisms decompose the incident white light into different colors on passing through the prism, i.e. they map the frequency to its wavenumber, when subjected to a spatial discontinuity and temporal dispersion.
In contrast, an inverted prism is a device that uses a temporal discontinuity and spatial dispersion  to map wavenumber to frequency. 

Another interesting demonstration of the effect of a time interface was demonstrated by Bacot \etal \cite{Bacot2016}, where the waves on the surface of a water bath generated by a source encounter a double time interface.
By providing a vertical jolt to the water bath the effective vertical acceleration was changed and this corresponds to an energy supply to the system. 
The double time interface is referred to as the \textquote{instantaneous time mirror}, which generates time reversed waves or backward propagating waves and forward propagating waves.
These time reversed waves converge to the shape of the original source, even when it is in a shape as complicated as that of the Eiffel tower.
Although there were existing techniques for time reversal  of waves, they were limited to either monochromatic (single frequency) waves or they were based on the use of emitters and antennas for recording, storing, and emitting time reversed waves. 
Temporal modulation achieves a new way of time-reversing broadband waves\cite{Bacot2016,tanter1998influence}. 


It is well known \cite{Lurie2006, Lurie2016} that when we have a wave propagating in a space-time metamaterial, then changing its microstructure properties generally requires addition of energy or removal of energy from the system.

A simple temporal laminate or space-time checkerboard constructed by temporally periodic switching of properties of a  spatial laminate has an exponential growth of energy in waves traveling through it for certain ranges of material parameters\cite{Lurie2006, lurie2007introduction, Lurie2016}.
Hence, depending on the type of modulation and the amplitude of the wave, this energy exchange can be of exponential form in time. 
This seriously limits the realization and application of space-time metamaterials. 
In some special geometries called \textquote{field pattern materials}, it was shown that one can achieve stable wave propagation without exponential growth\cite{mattei2017field}.
Recently, the growth of energy with time of edge waves in temporal laminates was shown by Movchan \etal\cite{movchan2022edge}.
Some lossy space-time metamaterials support wave propagation in special regimes, in which the waves can travel at constant amplitudes, where the exponential energy growth due to periodic time-modulation of properties is being compensated by energy dissipation\cite{torrent2018loss}.

The requirement of an energy source can be seen easily by imagining a chain of springs and masses with a wave propagating through it. 
The spring stiffness and/or masses change their values on encountering a time interface. 
In presence of the wave, if one wants to increase the stiffness of the springs at a certain time, then some amount of energy typically needs to be added to the system. 
Whereas, to make the springs more compliant (by reducing the stiffness) some amount of energy is typically released from the system. 
Similarly, switching the existing masses to lighter masses will typically release some kinetic energy, while switching to heavier masses will typically require adding energy to the system.

The space-time metamaterials considered in this work have properties changing discontinuously at time interfaces and we will refer to them as temporal metasurfaces.
The energy conserving scheme proposed in this section is applicable for temporal metasurfaces with waves propagating through them. 
We observe that certain material properties can be changed at specific instants of time without any energy cost.
Based on this observation we present our energy-conserving scheme below. 
In all the numerical simulations that follow, we implement a finite difference scheme to solve the equations of motion. 
All values are to be considered in SI units.
Length and time units in all figures are denoted by meters ($m$) and seconds ($s$) respectively.
We emphasize that the units are quite arbitrary and can be rescaled as one wants. The wave propagation speed will of course be altered if the spatial rescaling differs from the time rescaling, unless the mass units and spring constants are also rescaled accordingly.

Consider a discreet one dimensional (1-d) spring-mass chain with displacements inline with the one dimension.
Initially, each spring has the same stiffness $k$ and each mass has the same value $m$. 
Let $N$ be the number of springs and masses in the 1-d chain and $a$ be the lattice constant or length of the unit cell. 
A schematic representation of the 1-d unit cell is shown in the supplementary information in Figure \textcolor{red}{S1a}.
The displacements of the masses are denoted by $u_i(t)$ and are functions of time $t$.
For a wave propagating through this chain, the equation of motion for the $n^{th}$ mass is given by,
\begin{equation}\label{EOM one dimensional}
    \begin{split}
        m \Ddot{u}_n = - k (u_{n} - u_{n-1}) - k (u_{n} - u_{n+1})
    \end{split},
\end{equation}
where, the dot denotes partial derivative with respect to time $t$.

Suppose the $n^{th}$ spring is in equilibrium at a time instant $t_1$, i.e., there is zero stretching or compression of the spring, $\Delta u(t_1) = u_{n}(t_1) - u_{n-1}(t_1) = 0$.
At this instant, the potential energy of the spring is $\frac{1}{2}k\left(\Delta u(t_1)\right)^2 =0$, and the stiffness can be increased or decreased without adding or removing any energy from the spring-mass system. 
Alternatively, at some other instant of time say $t_2$, let the velocity $\Dot{u}_n(t)$ of the $n^{th}$ mass be zero, i.e., $\Dot{u}_n(t_2) = 0$.
Then at the instant $t_2$, the kinetic energy of this mass is given as,  $\frac{1}{2}m_n \Dot{u}_n^2(t_2)=0 $, and we can add or reduce this mass without any exchange of energy.
By considering only a single spring-mass system, one can easily see that changing the mass when it is at rest will not change the displacement amplitude but will change the period of oscillation.
Whereas, changing the stiffness of the spring at the equilibrium position will change both the amplitude and period of oscillation, but the velocity amplitude remains the same.
These serve as our basic mechanisms for time-modulation of material properties while conserving the total energy.
As the time modulation of properties depends on the specific instants of time, at which the spring is unstretched or the mass is at rest, this gives rise to the non-linearity in the problem. 
A numerical illustration of the energy conserving scheme in 1-d is shown in Figure \textcolor{red}{S2} of the supplementary information.
\begin{figure*}[htp!]
    \begin{subfigure}{0.48\linewidth}
        \centering
        \includegraphics[width=\linewidth]{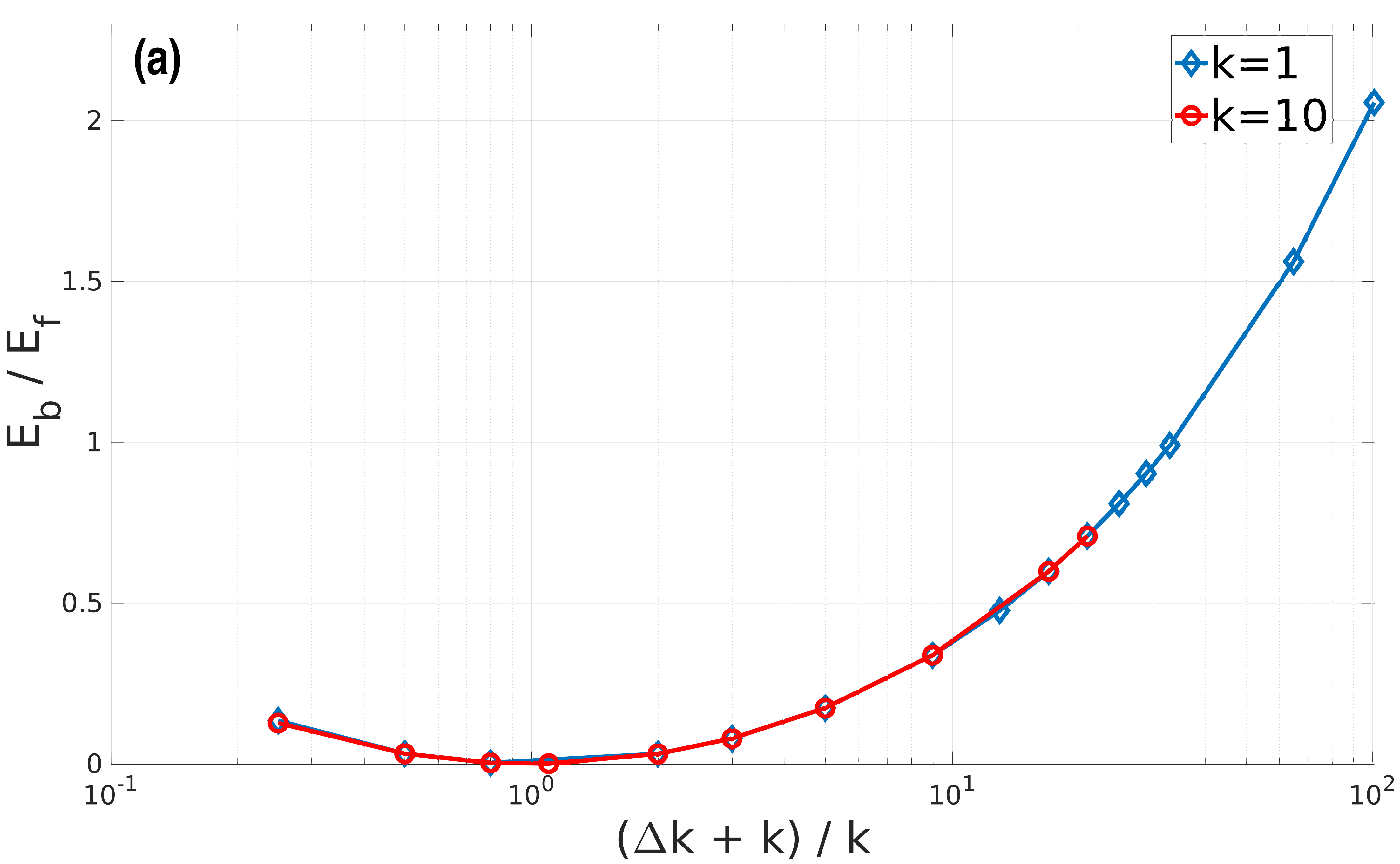} 
        \phantomsubcaption
        \label{fig:Energy ratio}
    \end{subfigure}
    \hfill
    \begin{subfigure}{0.5\linewidth}
        \centering
        \includegraphics[width=\linewidth]{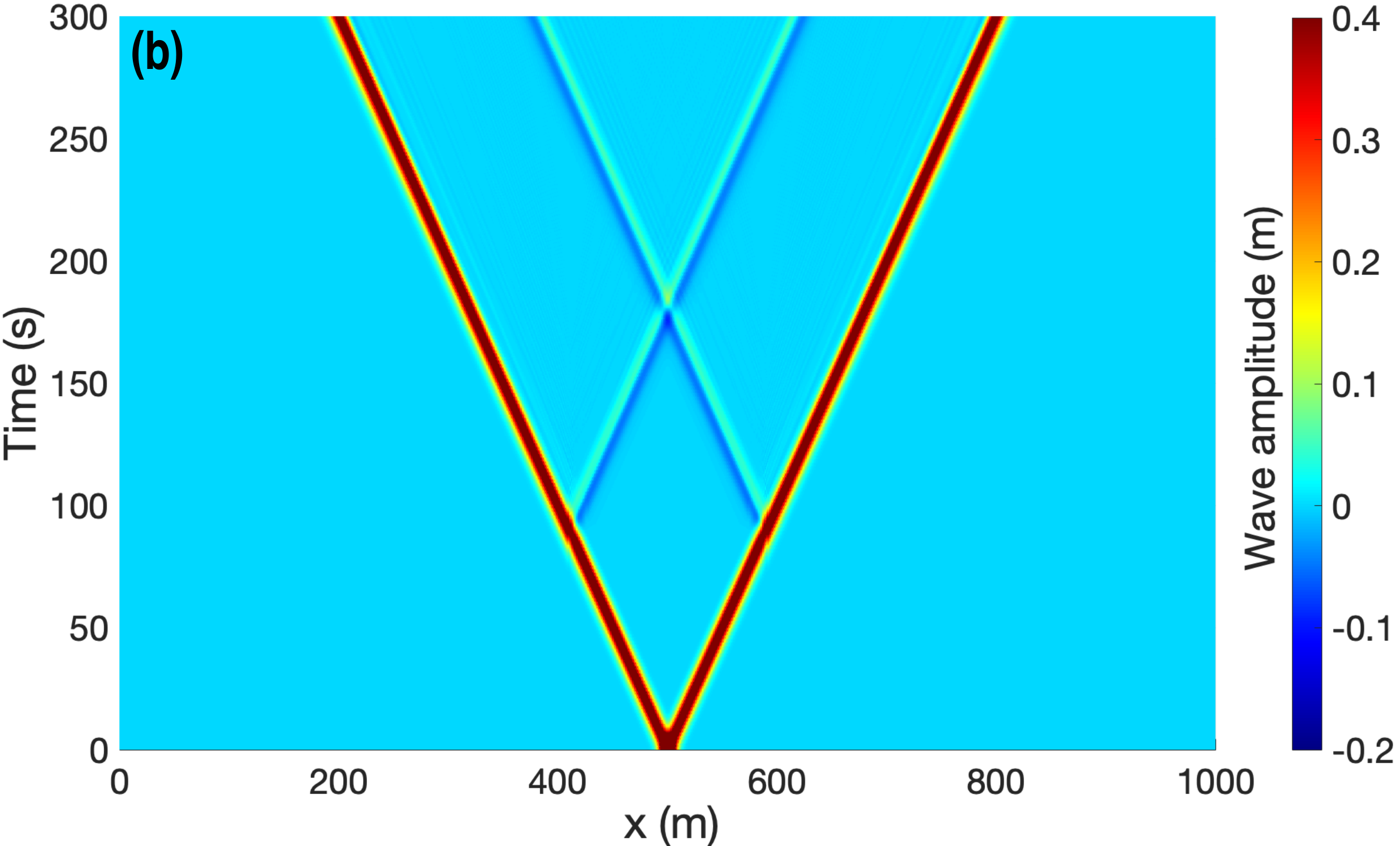}  
        \phantomsubcaption
        \label{fig:Regeneration_1d_stiffChange}
    \end{subfigure}
\caption{Time interfaces in 1-d.  
(a) Ratio of energies in the backward traveling wave ($E_b$) and forward traveling wave ($E_f$) vs. the change in stiffness  ($\Delta k$) using a single time interface with the abscissa in log scale. 
The blue (red) curve represents the 1-d spring mass chain with an initial spring stiffness value of $k=1$ ($k=10$). 
(b) Regeneration of the source in 1-d: The initial disturbance in the form of a Gaussian wave packet splits into two wave packets: traveling to right and left.
After the time interface we see each of the wave packets split into a forward traveling wave and a backward traveling wave. 
The backward traveling waves reconverge to form the source.}
\label{fig:Regeneration_1d}
\end{figure*}

The application of temporal metasurfaces to form time reversed waves discussed earlier is reconsidered here, but now using energy conserving temporal metasurfaces instead. 
In 1-d, the numerical calculations are presented using a discreet one-dimensional spring-mass system giving our energy conserving temporal metasurface.

Practically, changing the spring stiffness values is easier than changing the mass values. 
The examples below focus on changing the stiffness values at appropriate times near the time interface. 
A 1-d spring-mass chain with $k=1, m=1$ and $a=1$ as described before is considered here. 
The initial displacement for the masses is given by a Gaussian pulse of sufficient width to ensure  a homogenized behavior with negligible dispersion. 
The initial wave pulse splits into right and left traveling wave packets with negligible dispersion. 
A double time interface of width $\Delta t=3.2$ is generated after time $t=62.5$, where we change the stiffness of the springs by an amount $\Delta k = 10$ at the first time interface and then at the next time interface the stiffness is changed to the original value. 
The stiffness change at the time interface is carried out according to our energy conserving mechanism, i.e., the stiffness is changed whenever the springs are neither stretched nor compressed.  
Figure \ref{fig:Regeneration_1d_stiffChange} shows the space-time ($x-t$) plot of the displacement of masses as it encounters the double time interface. 
The left and right traveling waves hit the time interface and two sets of waves are generated from each of them; one  traveling backward in time as shown in blue, and the other traveling forward in time as shown in red. 
These time-reversed waves form the reconstructed source image. 

For a 1-d energy conserving temporal metasurface, Figure \ref{fig:Energy ratio} shows the ratio of energy in the backward propagating wave to energy in the forward propagating wave versus the change in stiffness at a single time interface as measured by the quantity $(\Delta k + k )/ k$. 
For high values of $(\Delta k + k ) /k$, the energy in the backward traveling waves ($E_b$) is seen to be higher than the energy in the forward traveling waves ($E_f$). 
The blue and red curves, representing the base stiffness values  of $k=1$ and $k=10$ respectively, lie on top of each other, thus indicating that the energy distribution is independent of base stiffness value.

\begin{figure*}[ht!]
    \centering
    \includegraphics[width=\textwidth]{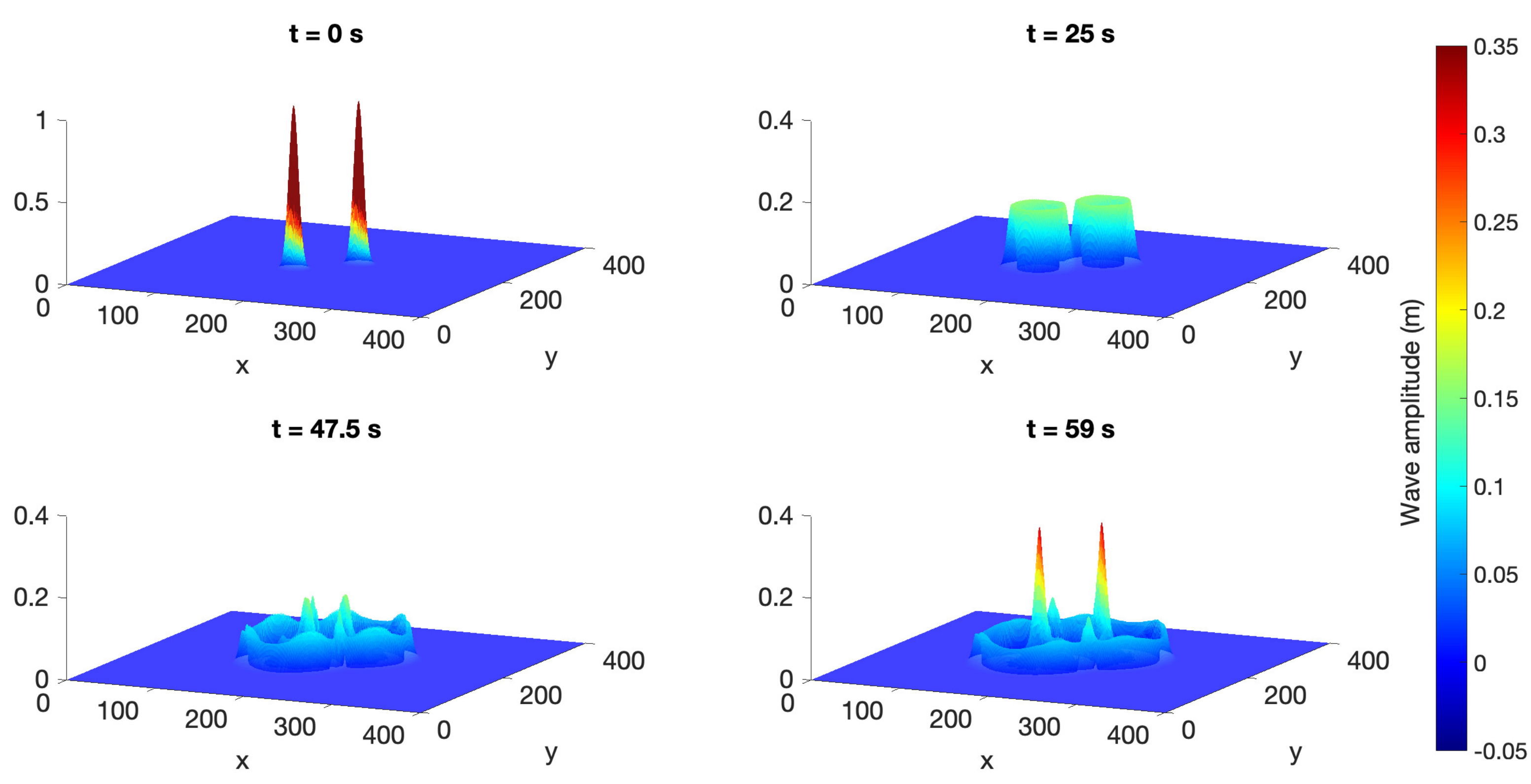}
    \caption{(Multimedia view) Reconstruction of 2 Gaussian waves: The initial disturbance or source is in the form of 2 Gaussian waves. 
    The first time interface occurs at time $t=25$.
    The time-reversed waves formed due to the double time interfaces re-converge to form the source at $t=59$.
    }
    \label{fig:Drops}
\end{figure*}
\begin{figure*}[ht!]
        \centering
        \includegraphics[width=0.98\textwidth]{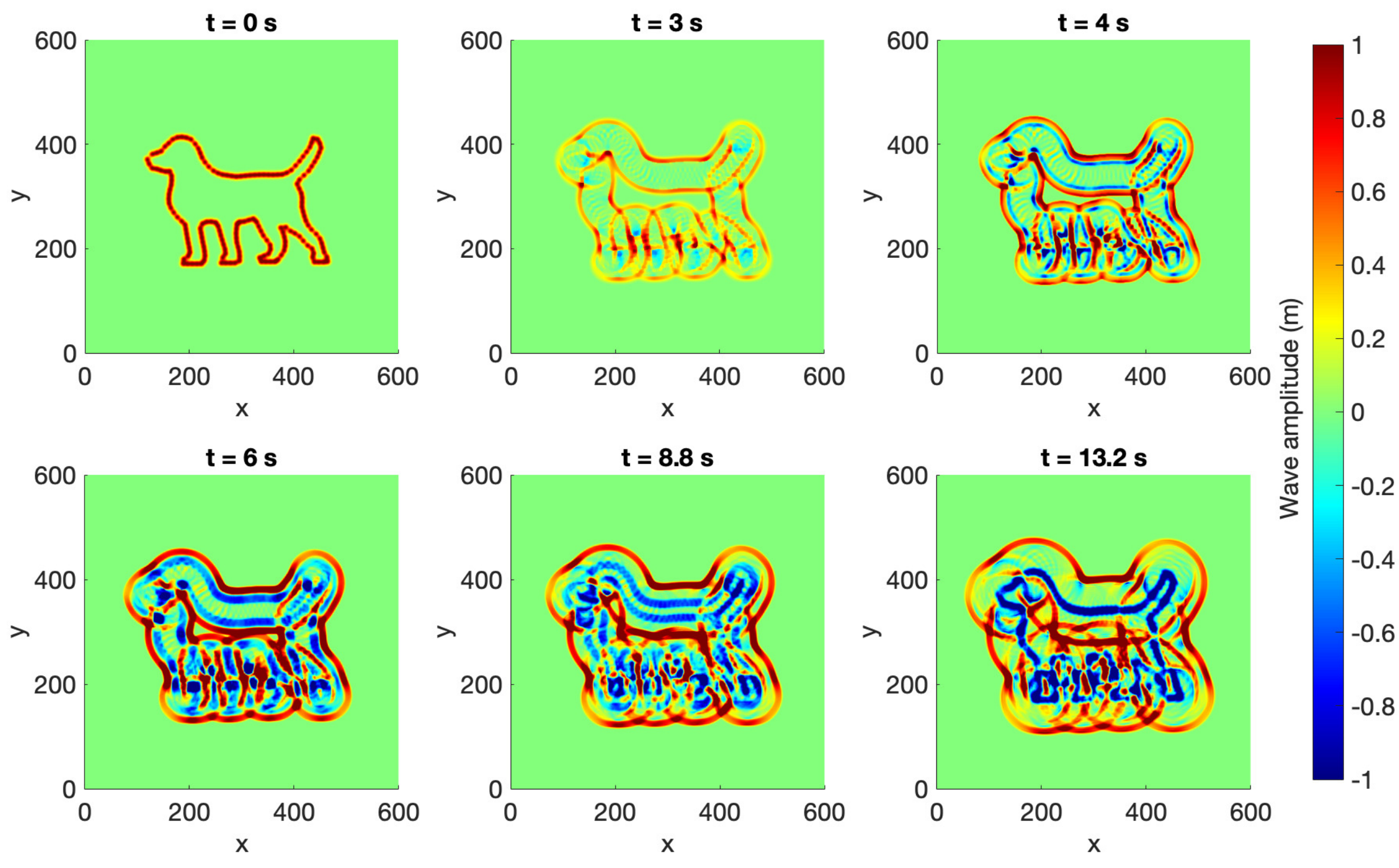}
        \caption{(Multimedia view) Dog: The initial source is in the shape of a \textit{dog}.  
        At the time interface ($t=3$), the stiffness is changed by an amount $\Delta k /k \approx -0.9$. 
        The time reversed waves seen in blue color due to the wave amplitude re-converge to form the source image.
        }
        \label{fig:Dog}
\end{figure*}

Next, examples of regeneration of sources in the form of  simple and complex shapes/images in 2-d are considered for temporal metasurfaces with energy conservation. 
For antiplane (shear) waves propagating in a three-dimensional (3-d) elastic material that is geometrically independent in one of the directions, say out-of-plane direction, we can model the shear restoring force using springs where the spring constants represent the shear modulus.
Hence, in our numerical simulations we effectively model the shear wave propagation in 3-d by a two-dimensional (2-d) spring-mass system with out-of-plane scalar displacements.
Such a 2-d spring-mass system, with the unit cell  as shown in supplementary Figure \textcolor{red}{S1b}, forms our energy conserving temporal metasurface.
We denote the number of unit cells in the $x$ and $y$ coordinate directions by $N_x$ and $N_y$ respectively. 
The mass in the $(i,j)^{th}$ unit cell is denoted by $m_{i,j}(t)$, its displacements denoted by $u_{i,j}(t)$, and  the stiffness values of 2 springs aligned along the $x$ and $y$ directions are denoted by $k_{x_{i,j}}(t)$ and $k_{y_{i,j}}(t)$ respectively.
This notation indicates that the material properties as well as the displacements vary both in space and time.
The equation governing the displacement of mass $m_{i,j}(t)$ is given by Equation \eqref{EoM 2d} and we drop the dependence on $t$ for conciseness of notation.
\begin{equation}\label{EoM 2d}
\begin{split}
     &\parderiv{}{t}\left(m_{i,j} \parderiv{{u}_{i,j}}{t}\right) = - k_{x_{i,j}} ({u}_{i,j} - {u}_{i-1,j} ) 
     \\
    &- k_{x_{i+1,j}} ({u}_{i,j} - {u}_{i-1,j} ) 
    - k_{y_{i,j}} ({u}_{i,j} - {u}_{i,j-1} ) 
    \\
    &- k_{y_{i,j+1}} ({u}_{i,j} - {u}_{i,j+1} ).
\end{split}
\end{equation}
The initial velocity of the masses is taken to be zero and the
2-d shape is prescribed as an initial out-of-plane (scalar) displacement for the masses. 
In the 2-d examples below, we change the stiffness of the springs at appropriate times near the time interface. 

A simple source taken in the form of two Gaussian functions is given by the expression $\exp{\left(-(\bfx - \bfmu_1)^2 / \sigma_1\right)} + \exp{\left(-(\bfx - \bfmu_2)^2 / \sigma_2\right)}$, with $\bfx=(x,y)$ representing the vector of $x$ and $y$ coordinates, $\bfmu_1 = (175,175)$, $\bfmu_2 =(225,225)$, and $\sigma_1=\sigma_2 = 50$. 
Figure \ref{fig:Drops} (Multimedia view) shows the snapshots of the displacement of masses (along the vertical axis) at subsequent instants of time. 
The initial displacement at time $t=0$, propagates as a wave and encounters the first time interface at $t=25$.
Near the first time interface, the stiffness of the springs is changed by an amount $\Delta k /k = 3$, and then at the second time interface ($t=25.5$) the stiffness is changed back to the original value. 
This generates the time-reversed waves (backward propagating waves) which converge to form the reconstructed source image  as seen from the snapshot at $t=59$.


\begin{figure*}[ht!]
    \centering
    \includegraphics[width=\textwidth]{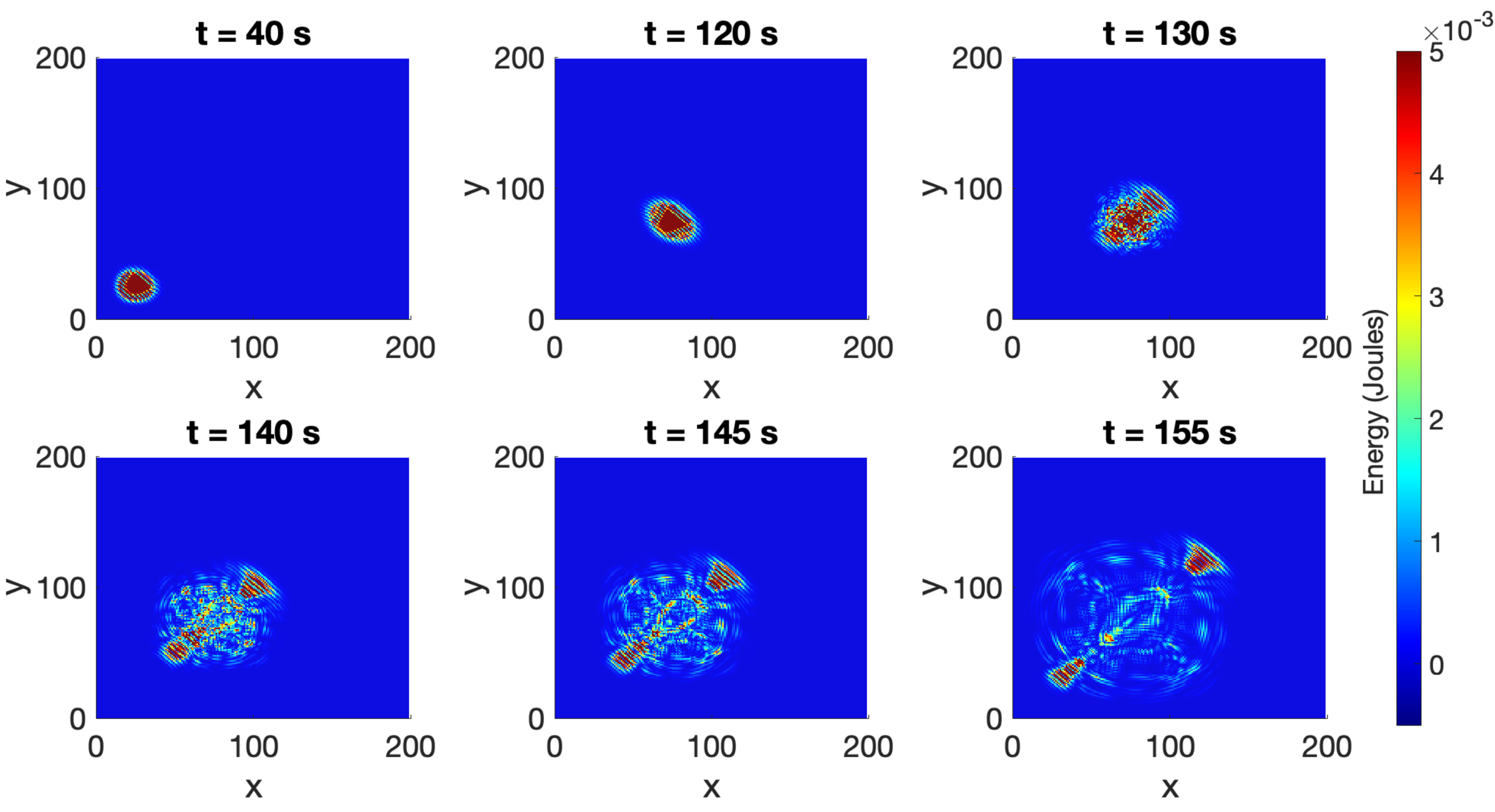}
    \caption{ Energy distribution resulting from scattering of a beam in a 2-d temporal metasurface:
    An incoming Gaussian wave pulse  encounters the temporal interface at $t=120$ and subsequent snapshots show that energy is distributed in all directions.
    }
    \label{fig:OneDropEnergy}
\end{figure*}
This next example shows that we can achieve source regeneration (\textit{Temporal image projection}) using an energy conserving temporal metasurface even when the source is in the form of a complicated image.
Figure \ref{fig:Dog} (Multimedia view) shows the evolution of the source in the form of an image of a \textit{dog}, into time-reversed waves and forward propagating waves. 
The spring stiffness is changed once by an amount $\Delta k / k \approx - 0.9 $ near the time interface. 
The initial displacement at time $t=0$  is allowed to evolve for some initial time until $t=3$. 
After it hits the single time interface, time-reversed waves are generated and due to the amplitude are seen in blue color at $t=4$, and eventually they converge to reconstruct the image of the dog at $t=13.2$.


In the above example, we observe that only a part of the outward propagating waves are time-reversed to reconstruct the image whereas the forward traveling waves result in an undesired interference which can be seen as the "noise" surrounding the reconstructed image.
It is desirable to have most of the energy from the incident wave concentrated in the backward propagating wave as compared to the forward propagating waves to get an accurate reconstruction of the source. 
While Figure \ref{fig:Energy ratio} shows that we can achieve this in the case of a single interface by increasing the amount of change ($\Delta k/k$ or $\Delta m /m$) in material  properties, it should also be noted that using a large value for $\Delta k/k$ will cause the wave front to be more spread out and one may get a blurred reconstructed image.
Similar results for source reconstruction obtained by changing the mass values are shown in  supplementary Figures \textcolor{red}{S4} and \textcolor{red}{S5}.

Temporal modulations in an energy conserving temporal metasurface occur at times neighboring the time interface, resulting in the material being anisotropic and inhomogeneous for those times.
Thus, a Gaussian beam propagating in a certain direction in an energy conserving temporal metasurface after encountering a time interface has the total energy  scattered differently in all directions. 

A Gaussian beam propagating in a 3-d material is effectively modeled by considering a Gaussian pulse propagating along a line in a 2-d spring-mass system.
Figure \ref{fig:OneDropEnergy} shows the total energy distribution in the spring-mass system resulting from scattering of the incident wave due to the temporal interface. 
An incoming Gaussian pulse from the bottom left corner encounters a temporal interface at $t=120$  after which the stiffness is changed once by $\Delta k/k=3$ at appropriate times.
In the subsequent snapshots we observe that, while most of energy is distributed  along the line making an angle of $45^\circ$ to the x-axis (seen in dark red), some amount of energy is scattered in every other direction.
This illustrates the inhomogeneous nature of the temporal metasurface at times neighboring the temporal interface.


In conclusion, we have proposed an energy conserving mechanism for temporal metasurfaces caused by time modulation of material properties.
The mechanism is demonstrated using a discreet spring-mass system in one and two dimensions, with an application to reconstruction of the source image using time-reversed waves. 
For this particular application, the operation is restricted to the homogenized regime to get a better reconstructed image of the source. However, the energy conserving mechanism itself is not restricted to the homogenized regime. 
In practice, one can change the stiffness of the spring by attaching a spring in parallel to it between the same points (i.e. a spring of same length) when the spring is neither stretched nor compressed. 
Even if one cannot locate the perfect instant to carry out this operation, the energy exchange can always be bounded between some limits.
With some caution, energy conserving temporal metasurfaces can be realized easily using electrical transmission line structures and a brief discussion on this topic can be found in the supplementary information. 
The energy conserving mechanism can also be extended to interfaces at an angle in space-time and to some complicated geometries in space-time.
For waves propagating in continuum media in 1-d, the modulus can be changed at an instant of time  when the stress is close to $0$, and the result can be easily generalized to two and three dimensions.
Thus, the energy conserving mechanism can be extended to continuum systems as well.
Experimental realization of space-time metamaterials (with more than one temporal interfaces) having close to sharp temporal interfaces that was previously not possible due to the energy exchange requirements, now seems possible with the proposed scheme. 

See supplementary information document for additional details related to this work.

The authors are grateful to the National Science Foundation for support through grant DMS-2107926. 

 
\section*{Author Declarations}
\subsection*{Conflict of Interest}
The authors have no conflicts to disclose.

\subsection*{Authors' Contribution}
G. W. Milton suggested the problem and K. J. Deshmukh did all other work.
\section*{Data Availability}
Aside from the data available within the article, any other data are available from the corresponding author upon request. 

\section*{References}

\bibliography{main}
\end{document}